\documentclass{article}
\usepackage{spconf,amsmath,graphicx}
\usepackage{epstopdf}

\title{Performance Advantages of Deep Neural Networks For Angle of Arrival Estimation}
\name{Oded Bialer $^1$, Noa Garnett $^1$, and Tom Tirer $^{1,2}$} 
\address{$^1$ General Motors - Advanced Technical Center Israel \\$^2$ School of Electrical Engineering, Tel Aviv University, Tel Aviv, Israel}
\begin{document}
%
\maketitle
\begin{abstract}
The problem of estimating the number of sources and their angles of arrival from a single antenna array observation has been an active area of research in the signal processing community for the last few decades. When the number of sources is large, the maximum likelihood estimator is intractable due to its very high complexity, and therefore alternative signal processing methods have been developed with some performance loss. In this paper, we apply a deep neural network (DNN) approach to the problem and analyze its advantages with respect to signal processing algorithms. We show that an appropriate designed network can attain the maximum likelihood performance with feasible complexity and outperform other feasible signal processing estimation methods over various signal to noise ratios and array response inaccuracies. 
\end{abstract}
\begin{keywords}
Angle of arrival, deep neural networks, model order determination, single snapshot 
\end{keywords}
\section{Introduction}
\label{INTRO}

Estimating the angles of arrival (AOAs) of an unknown number of multiple sources transmitting unknown signals is a challenging problem with
various applications such as  localization, radar, sonar, anti-jamming and wireless communication, and has
been an active area of research in the signal processing community for the last few decades \cite{Hamid}.

The maximum likelihood (ML) AOA estimator for a single source is the angle of the Bartlett beamformer peak \cite{VAN_TREES}.
In case the number of sources is unknown, an initial step of model order determination is required in order to formulate the ML-AOA estimator \cite{VAN_TREES}.
The model order can be estimated by MDL and AIC methods \cite{WAX_detect}. However, both require multiple realizations (snapshots).
Furthermore, even for a known number of sources, the computational complexity of ML estimation grows exponentially with this number. When the number of sources is more than two and the range of possible angles is wide (large search grid) the ML complexity is extremely high and hence it is impractical for implementation.

Applying the Bartlett beamforming for estimating multiple angles of arrival results in significantly inferior performance compared to ML since it suffers from a large bias and even fails to resolve sources when their angles are closer than the beamforming 3 dB width. 
Various iterative approximate ML estimators have been proposed with reduced complexity, such as alternating projections (AP) \cite{AP}, IQML \cite{IQML}, EM \cite{EM}, and Orthogonal Matching Pursuit (OMP) \cite{OMP_REF}. Super-resolution methods, such as MUSIC \cite{MUSIC_REF}, ESPRIT \cite{ESPRIT}, and MVDR \cite{CAPON}, have been also applied for AOAs estimation. These methods rely on having multiple independent realizations of sources. However, in many cases there is only a single realization from which the AOAs need to be estimated, for example when sources and/or receivers are moving. In the special case of uniform linear arrays (ULAs), it is possible to create virtual realizations using spatial smoothing \cite{SpatialSmooting}. Yet, this technique is effective mainly in high signal to noise ration (SNR), and also inherently shortens the array aperture, which in turn reduces its angular resolution and accuracy.

Estimating the AOAs with machine learning methods is an alternative to signal processing methods. 
The pioneering works \cite{S1}-\cite{S3} demonstrated DOA estimation with Hopfield neural networks. These were networks with a single hidden layer and relatively small number of neurons. Southall et. al. \cite{S4} applied a larger neural network with three hidden layers to estimate the DOA of a single source and showed a performance advantage with respect to DOA estimation with monopulse. 
Other more recent works \cite{D1}-\cite{D4} have applied deep neural networks (DNNs) for 
estimating acoustic sources direction/position from a large number of realizations of microphones array. It was shown that neural networks attain relatively good accuracy compared to MUSIC \cite{MUSIC_REF} in challenging acoustic room environment conditions, such as reverberations and high noise.  

In this paper, we apply a DNN to estimate the number of sources and their angles of arrival based on a {\em single} realization, and analyze the DNN advantages with respect to various signal processing methods, including the ML estimator and its approximation by the prominent AP algorithm \cite{AP}. 
The first contribution of this paper is to show that in the case of a known number of multiple sources the DNN is a practical AOAs estimator that can attain the intractable ML estimator performance, and outperforms other leading practical signal processing methods. The second contribution is to show that for single snapshot the accuracy of the DNN model order estimation is significantly better than the classical MDL and AIC methods, that require the use of spatial smoothing in this case.

\section{System model}\label{SYS_MODEL}
We consider an array of $N$ antenna elements linearly spaced that are receiving signals from multiple point sources at different angles.
It is assumed that these sources have relatively narrow band and are at far-field, thus the received array signal can be expressed by
\begin{equation} \label{SYS_MODEL_1}
\boldsymbol{y}=\sum _{m=0}^{M-1}\boldsymbol{a}(\theta_m)s_m+\boldsymbol{v},
\end{equation}
where $\boldsymbol{v}$ is the noise vector, $M$ is the number of sources (number of angles of arrival), $m$ is the source index, $s_m$ and $\theta_m$ are the unknown deterministic complex signal coefficient and the angle of the $m$-th source, respectively, and
\begin{equation} \label{SYS_MODEL_2}
\boldsymbol{a}(\theta)=\left[\begin{array}{c} {e^{\frac{j2\pi}{\lambda}x_1sin(\theta)}}\\ {:} \\ {e^{\frac{j2\pi}{\lambda}x_{N}sin(\theta)}}\end{array}\right],
\end{equation}
is a steering vector for an angle of arrival $\theta$, where $\lambda$ is the wavelength, and $x_n$ is the $n$-th antenna position with respect to the linear array center point.
The number of sources, $M$, and their angles $\theta_0,..,\theta_{M-1}$ are unknown and are estimated based on a single array realization (snapshot) with a neural network as described in the next section.

\section{DNN Multi-Source AOA Estimation}\label{DNN_ARCHETICTURE}

\begin{figure}[t]
\centering
\includegraphics[width=8cm]{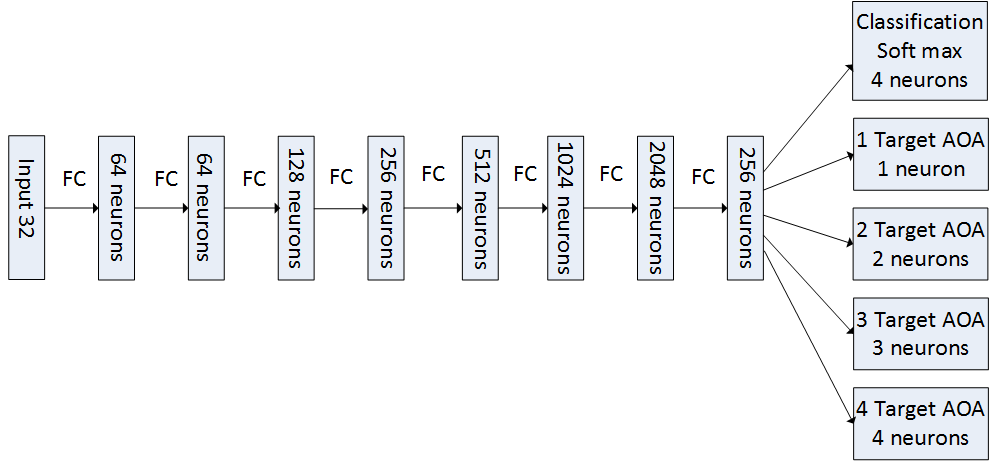}
\caption{\label{NET_ARCH}The neural network architecture with 8 hidden fully connected (FC) layers, followed by classification (softmax + cross entropy loss) of the number of sources (1-4), and AOAs regression for each one of the classes.}
\end{figure}

The DNN architecture that was used to estimate the number of sources and their angles-of-arrival is depicted in Fig.\ref{NET_ARCH}. The DNN was designed for an array of 16 antennas ($N=16$) uniformly spaced with $\lambda/2$ spacing, and for estimating up to 4 sources ($M\le4$).
It is straight forward to modify the same DNN architecture for other system configurations, such as non-uniform spacing, different number of antennas, and different number of possible sources.

The input to the network is the array received signal of $2N$ values, which are the real and imaginary components of the array response. Following the input layer there
are 8 fully connected layers, each followed by batch normalization \cite{batchNorm} and ReLU nonlinearity. The number of neurons in each layer is shown in Fig.\ref{NET_ARCH}.
The network has 14 outputs, which are related to five different groups, as described in the following.

The first group is the classification of the number of sources, and includes four outputs, each representing the probability of the number of sources between 1-4. The classification output decision is the index of the maximal probability output. The loss function of the classification was the cross entropy after softmax. The other four groups of outputs are the angles of arrival estimations for each of the four classification options. The angle estimation loss function for each of these groups was the squared root of the mean square error (RMSE) between the true sources angles and their closest estimation, which is mathematically expresses by
\begin{equation} \label{DNN_1}
RMSE = \sqrt{E\Biggl\{\frac{1}{M}\sum_{m=0}^{M} \underset{k}{\mathrm{min}} \bigl\{(\hat{\theta}_k-\theta_m)^2\bigr\}\Biggr\}},
\end{equation}
where $\theta_m$, and $\hat{\theta}_k$ are the true and estimated angle of arrivals, respectively, and $k\in{0,1..M-1}$. During training, for each example only the RMSE of the true number of classes was considered.

The network parameters were trained and tested using simulated data according to the system model described in Section \ref{SYS_MODEL}. The network was trained with stochastic gradient descent with momentum, where each descent iteration included a batch of 4000 realizations from different number of sources and different SNR values, and also different signal coefficients, $s_m$.

The network described above attained best performance with respect to other architectures that were tested as well, including a convolutional network (ConvNet), and a fully connected network with smaller number of layers or smaller number of neurons per layer.
A fully connected network with a dense net architecture \cite{DENS_NET} was also tested, and showed comparable results to the network in Fig.\ref{NET_ARCH}.
The performance comparison of different network architectures is presented in Section \ref{RES_SEC}.


\subsection{Complexity analysis}
As mentioned in Section \ref{INTRO}, a major practical limitation of the ML estimator of multiple sources AOAs (formulation can be found in \cite{VAN_TREES}) is its very high computational complexity. Therefore, we evaluate next the complexity of the proposed DNN estimator and compare it to the complexity of the ML estimator and also to the AP estimator \cite{AP}, which as described in Section \ref{INTRO} is a low complexity approximation of the ML.

During the deployment stage (after learning the network parameters), the complexity of the DNN in Fig.\ref{NET_ARCH} is given by the number of neurons, which is mainly governed by the layer with 2048 neurons, hence the complexity of the DNN is given by $O(2048^2)=O(4.19^6)$. The complexity of the ML estimator 
is $O((N^2+M^3)*P^M)$, and the complexity of the AP estimator is $O(N^2*P*K)$, where $P$ is the number of angle search grid points, and $K$ is the number of AP iterations.

Let us consider the following example for comparison. For a 50 degrees filed-of-view and a grid spacing of 0.1 degree, we have that P=50/0.1=500, and the complexity of the ML estimator is $3.2^{13}$. For the same case, the complexity of the AP with 10 iterations (which were necessary to get to good performance as shown in Section \ref{RES_SEC}) is $O(10^6)$. It is realized that the DNN has similar complexity to the AP and both methods have 7 order of magnitude lower complexity than the ML estimator. The significant lower complexity makes both DNN and AP much more practical estimators than the ML.
We also note that the DNN is much faster than AP when executed on GPUs.

\section{Results and Discussion}\label{RES_SEC}
\begin{figure}[t]
\centering
\includegraphics[width=5.5cm]{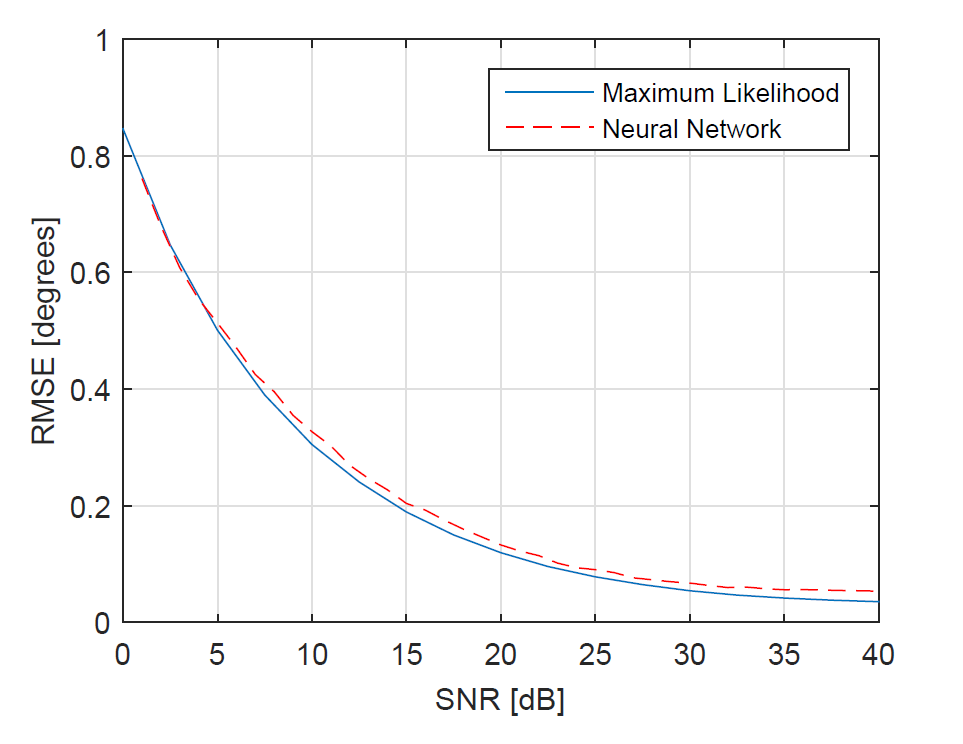}
\caption{\label{RES_2_TARGETS}DNN vs. maximum likelihood AOA estimation performance of two sources in the FOV of 20 degrees}
\vspace{3mm}
\includegraphics[width=5.1cm]{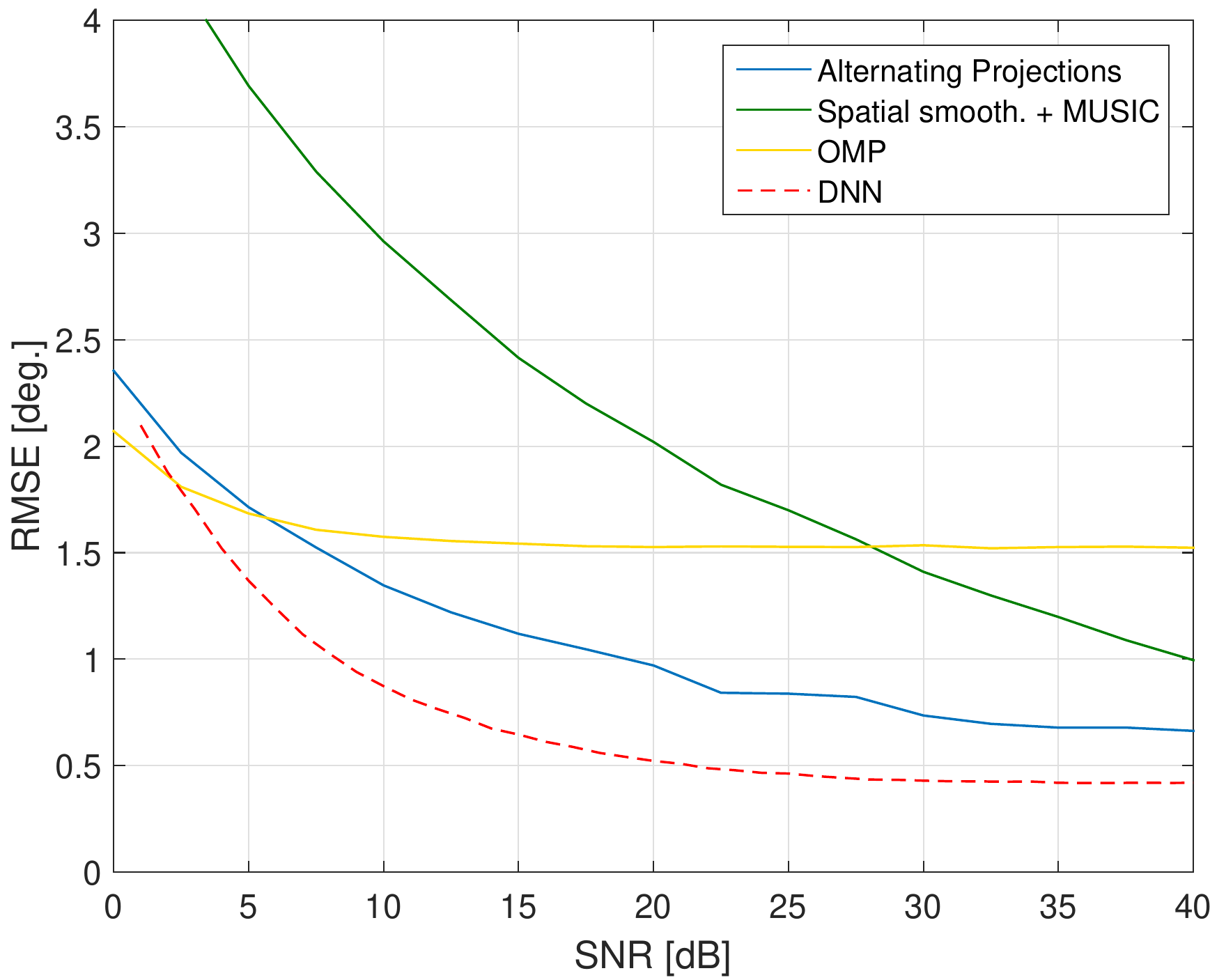}
\caption{\label{RES_4_TARGETS_EST}AOA estimation performance for four sources in the FOV of 50 degrees}
\end{figure}

We first compared the AOAs estimation of the DNN described in Section \ref{DNN_ARCHETICTURE} with the ML estimator for two sources with equal gains from a single array realization (snapshot). 
In this case the ML estimator still has feasible complexity and hence can be tested in a reasonable runtime. Fig. \ref{RES_2_TARGETS} presents the RMSE results of both estimators, where the RMSE calculation is given in \eqref{DNN_1}, with $M=2$. The results show that the DNN AOA estimator attains the RMSE performance of the ML over a broad range of SNR values. Asymptotically, in high SNR, the ML attains the CRLB (lower bound) on the AOA estimation RMSE and hence the DNN AOA estimator also attains the ultimate RMSE performance.

Next, we tested the performance of the DNN AOAs estimation of four sources from a single array realization. In this case, the number of sources was known, but their angles of arrival were unknown and uniformly distributed within a filed of view (FOV) of 50 degrees. The complex coefficients of each source, $s_m$, in \eqref{SYS_MODEL_1} had random phase uniformly distributed in the range of $[0,2\pi]$, and random amplitudes uniformly distributed in the range of $[0.5,1.5]$. In the case of four sources the complexity of the ML estimator is too high, and its runtime is unfeasible. Hence we compare the RMSE of the DNN AOA estimation to three prominent signal processing algorithms: AP \cite{AP}, MUSIC \cite{MUSIC_REF}, and OMP \cite{OMP_REF}. For MUSIC, in the absence of multiple array realizations, we applied first a spatial smoothing step \cite{SpatialSmooting}, where the 16 element ULA were partitioned into 9 overlapping sub-arrays (equivalent to 9 snapshots), each consisted of 8 elements. This choice was obtained empirically, for best performance.
The results are presented in Fig.~\ref{RES_4_TARGETS_EST}.
The DNN has a performance advantage over all reference methods.
It reaches an RMSE less than 0.5 degrees which is very low considering the relatively large 3dB beamwidth of the array, which is 10 degrees.

\begin{figure}[h]
\centering
\includegraphics[width=6cm]{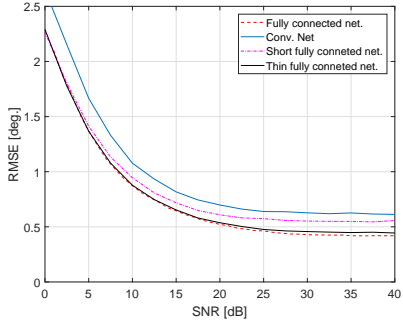}
\caption{\label{network_comparison}AOA estimation performance of different network architectures for four sources.}
\end{figure}

We have also tested the performance of the proposed fully connected (FC) DNN in Fig.~\ref{NET_ARCH} with three other networks: a ConvNet with 8 layers +  FC layer, a shorter FC network with only 6 hidden layers, and a thinner FC network with 8 hidden layers and smaller number of channels (obtained by limiting the number of channels per layer to 400). The results presented in Fig.~\ref{network_comparison} show that the network in Fig.~\ref{NET_ARCH} attains best performance. The ConvNet has noticeable degraded performance compared to the fully connected network. Furthermore, shortening the depth of the network causes some degradation, while reducing the number of channels results in minor degradation. We have also tested a fully connected network in a dense net architecture \cite{DENS_NET}, where each layer receives the accumulated inputs of all previous layers, and observed similar performance as the network in Fig.~\ref{NET_ARCH}.

\begin{figure}[t]
\centering
\includegraphics[width=4.9cm]{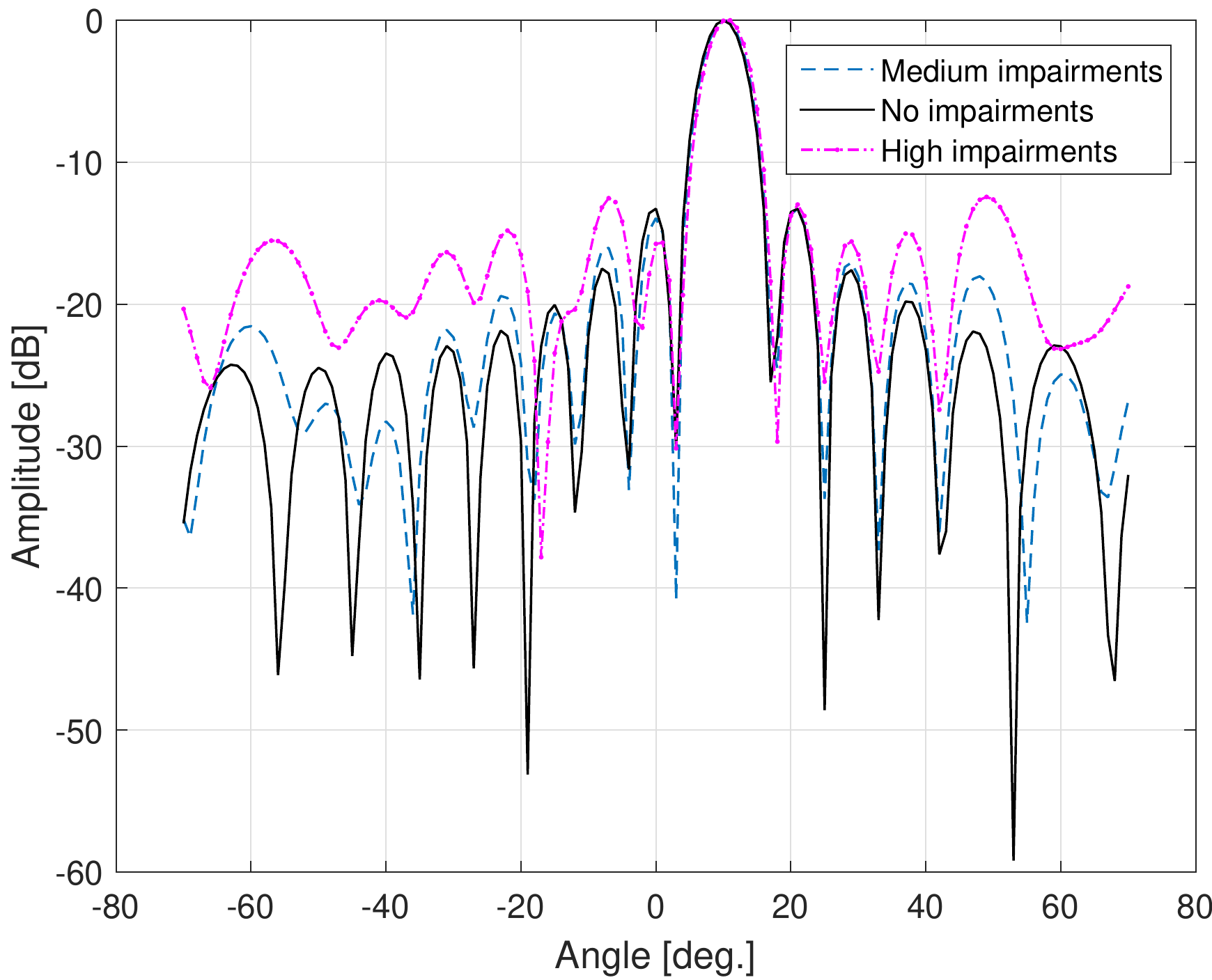}
\caption{\label{Spect_impairments}Beampattern (Bartlett beamformer spectrum) for different levels of impairments}
\vspace{2mm}
\includegraphics[width=4.9cm]{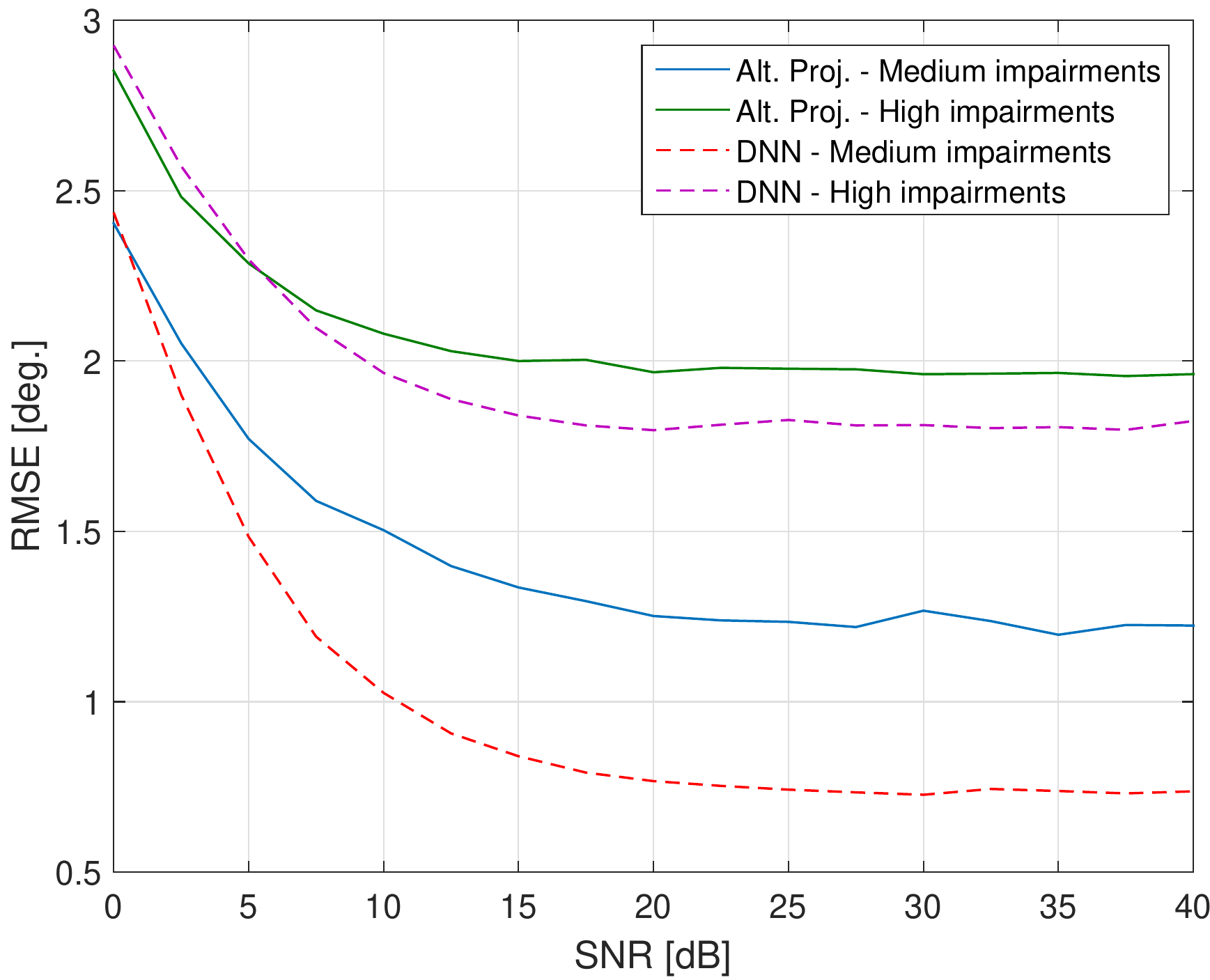}
\caption{\label{RES_impairments}AOA estimation performance for four sources in the FOV of 50 degrees, for different levels of impairments}
\vspace{4mm}
\includegraphics[width=4.9cm]{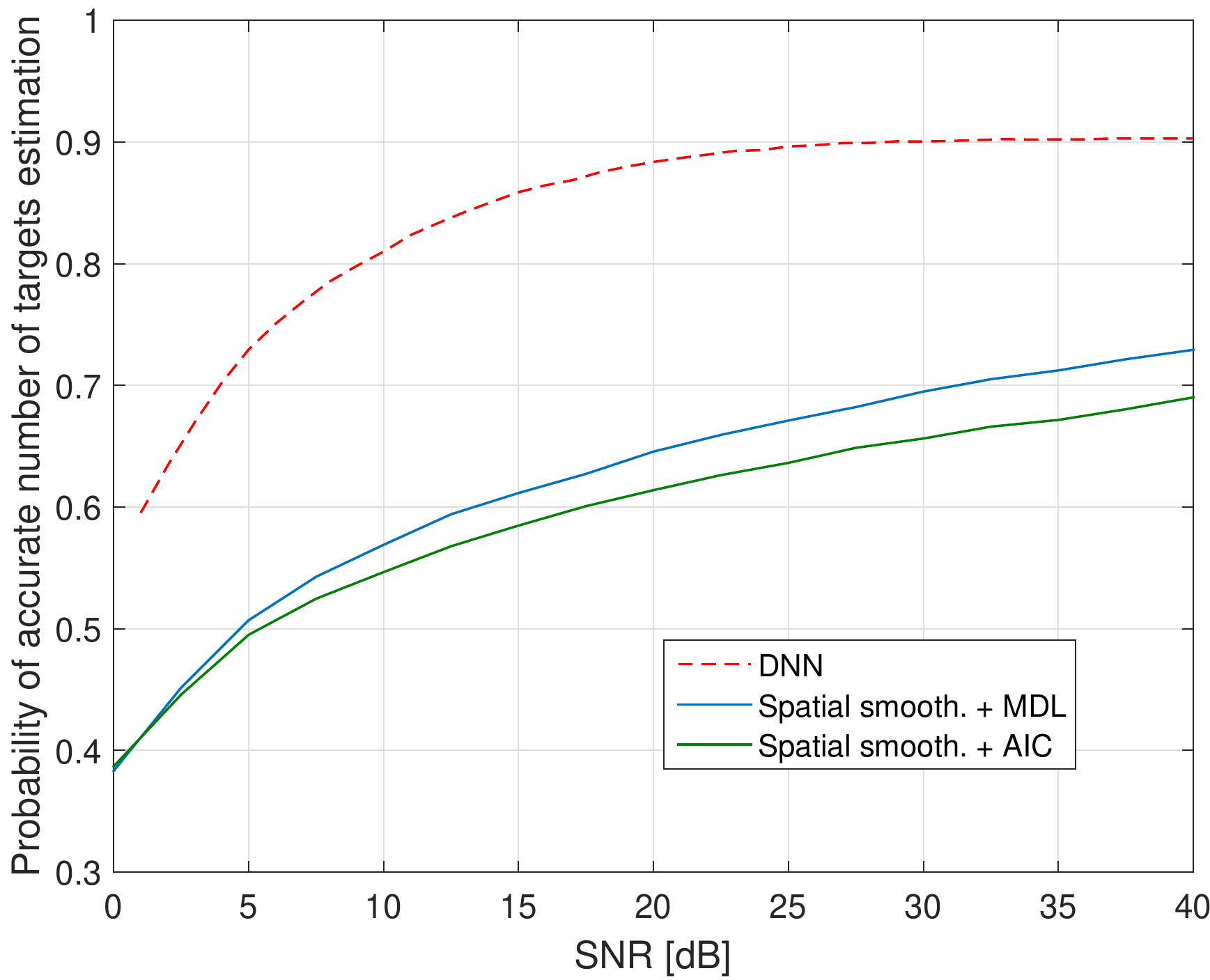}
\caption{\label{RES_4_TARGETS_CALSS}Number of sources estimation performance}
\end{figure}

Next, we examined the robustness of the DNN to model errors, which often occur in practical systems.
The model errors were obtained by feeding the DNN the vector $\boldsymbol{H}\boldsymbol{y}$ instead of $\boldsymbol{y}$, where $\boldsymbol{H}$ is a mismatch matrix that is unknown to the estimator. The diagonal of $\boldsymbol{H}$ had unit amplitude and Gaussian phase offsets with standard deviation denoted by $\sigma_{\theta}$, representing phase offsets between the antenna channels. The off-diagonal elements of $\boldsymbol{H}$ modeled the cross talk between the channels by having complex Gaussian random variables with an amplitude attenuation factor of $\Gamma$ below unity (below the diagonal amplitude). We have tested two levels of impairment: medium impairments with $\sigma_{\theta}=5^\circ$, $\Gamma=30dB$ and high impairments where $\sigma_{\theta}=10^\circ$, and $\Gamma=20dB$. The effect of these impairments on the array response is demonstrated
in Fig. \ref{Spect_impairments} by comparing the array Bartlett beamformer for the case with and without impairments. The DNN and AP AOAs estimation performance for four sources with two levels of impairments is shown in Fig.~\ref{RES_impairments}. As expected, the performance of both the DNN and AP is degraded compared to Fig.~\ref{RES_4_TARGETS_EST}. However, in the case of high mismatch the DNNs performance advantage over AP is even larger than the case without mismatch. Meaning that the DNN is relatively robust to model errors, although it was not trained with model errors. A possible explanation is that the DNN was trained with effective regularization due to the very large number of diverse examples with different AOAs and various SNR values.

Finally, we tested the performance of the DNN number of sources estimation from a single array realization (classification performance). In this test, for each received signal the number of sources was randomly chosen from a uniform distribution between 1-4. We compared the DNN with two prominent signal processing alternatives, MDL and AIC \cite{WAX_detect}, and limited their results to 1-4. These methods require multiple realizations, hence we applied an initial spatial smoothing step that generated 9 virtual realizations, as mentioned above for the implementation of MUSIC. Fig. \ref{RES_4_TARGETS_CALSS} presents the probability of accurate estimation of the number of sources vs. the SNR. It is realized that the DNN accuracy is significantly better than the reference methods. The DNN reaches 0.9 probability of accuracy in high SNR, while the best reference method attains only 0.72. We have observed that most of the DNN classification errors occurred when two or more of the true source angles of arrival were randomly chosen to be very close. In this case the classifier falsely
detected the close sources as a single source, and as a result the number of estimated sources was smaller than the true value.

\section{Conclusions}\label{CONCLUSION_SEC}
We have shown that the DNN approach for estimating the number of sources and their AOAs from a single realization has several advantages over signal processing methods. For AOA estimation, the DNN reaches ML performance with significantly lower complexity and unlike the ML it is feasible for implementation even when the number of sources is large. Furthermore, the DNN also outperforms leading signal processing methods with feasible complexity (alternatives to the ML), such as AP, MUSIC, and OMP.
For model order estimation (from a single snapshot), the DNN significantly outperforms the prominent MDL and AIC methods.
It was also shown that the DNN is relatively robust to imperfectness (mismatch) in the array response.

\vfill\pagebreak



\end{document}